# Characteristics of the hall current surrounding a magnetized filamentary plasma

A. Tonegawa, T. Okawara, K. Kawamura

*Department of Physics, School of Science, Tokai University*
*1117Kitakaname, Hiratsuka, Kanagawa 259-1292, Japan*

The hall current in the periphery of a filamentary plasma with the inwardly directed radial electric field, $E_r$ has been investigated experimentally by utilizing a TPD type dc discharge. When the cusp magnetic filed $B_\theta$ is applied perpendicular to the stationary background magnetic field $B_z$, the hall current by the $E_r \times B_\theta$ drift of electrons is induced in the periphery of the magnetized filamentary plasma.

1. Introduction

The current flow on the periphery of the magnetized plasma is though to be a key issue in magnetic reconnection and confinement of the nuclear fusion. In boundary region of a steady-state magnetized plasma, the density gradient-scale length $L_n$ of the plasma column is smaller than the ion Larmor radius $\rho_i$ and much larger than the electron Larmor radius $\rho_e$, ($\rho_e \ll L_n < \rho_i$). This situation can be viewed that ions aren't trapped in a radial electric-potential-well which is a result of the charge separation between an electron "string" confined along an axial magnetic field. Because the ions can be considered unmagnetized under this condition, a axial current can arise in our plasma due to the $E_r \times B_\theta$ drift motion of the electrons. If the multi-cusp magnetic filed $B_\theta$ is applied perpendicular to the stationary background magnetic field $B_z$, the current flow causes by the relative motion of electron and ions, that is, the hall current.

This paper presents the preliminary results of the hall current in the periphery of a filamentary plasma [1,2]. When the multi-cusp magnetic filed $B_\theta$ is applied perpendicular to the stationary background magnetic field $B_z$ and the electric filed in the radial direction $E_r$, the hall current by the $E_r \times B_\theta$ drift of electrons is induced surrounding a magnetized filamentary plasma. This hall current generates the magnetic field in an outside area of the plasma column resulting the inner magnetic field in the plasma column decrease and the outer magnetic field increase. The characteristics of the hall current indicates that the increase of the diamagnetic effects in a plasma.

2. Experimental apparatus

Our experiment is performed by using the string plasma device (SPD) (see Fig.1) developed at Tokai University [1, 2]. The device is divided into two regions: the discharge region at the argon neutral pressure of 70 Pa and the low-pressure ($7 \times 10^{-3}$ Pa) experimental region. Two regions are differentially pumped through a narrow path installed by cylindrical floating electrodes and an anode. The plasma is continually generated inside the discharge region by means of a dc discharge between the anode and the electron-emitting $LaB_6$ hot cathode. The cathode is placed in the field-free region of the cusp magnetic field in order to minimize the damage of the cathode surface due to ion bombardments.

The plasma is compressed radially by the converging magnetic field and the inwardly directed radial electric field formed through a series of tapered floating electrodes. The formation of this radial electric field is expected by the positively charged-up floating electrodes. This is supported by the fact that ions can reach the electrode walls much faster than electrons since the Larmor radius of ions is much larger than that of electrons. The diameter of the plasma passing through the narrow path is reduced from 80 to 2 mm. Through the 1-mm-diam anode hole, the plasma is introduced into the low neutral-pressure experimental region along the uniform magnetic field $B_z$. The strength of the axial magnetic filed in the experimental region is 0.19 T. A metal target (60 mm in diameter) which terminates the plasma flow is placed 250 mm from the anode hole along the z axis.

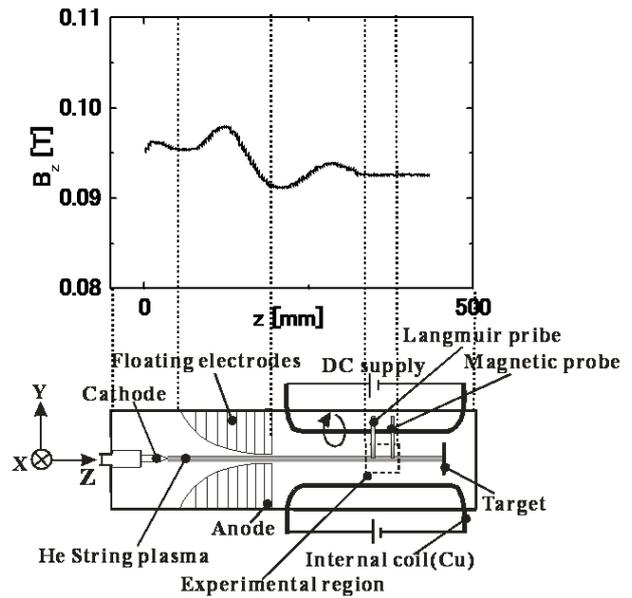

In our plasma, the effective plasma radius, $r_p$, under the present conditions is within a range of a few millimeters, which is much larger than the electron Larmor radius but is smaller than the ion Larmor radius. The plasma space potential $V_p$ is high in the outer region of the plasma ($r>r_p$) and is minimum at the center. Consequently, the inwardly directed radial electric field, $E_r$, naturally sets up and it also varies in the range of $r_p$. This is a cause for the formation of the well-type radial potential profile in our plasma.

The multi-cusp magnetic filed $B_\theta$ is applied perpendicular to the stationary background magnetic field $B_z$. The magnetic filed $B_\theta$ is established by pulsing axial current ($I_{in}$= 130 A, 5 Hz) through four parallel internal coil. The plasma space potential, $V_p$, the electron density, $n_e$, and the electron temperature, $T_e$, are measured by a small one-side disk Langmuir probe. The probe tip (tungsten, 0.5 mm in diameter) is embedded in a 1-mm-diam ceramic tube. The plasma space potential $V_p$ is determined from a peak position of the $dI_p/dV_p$ curve of the $I_p$-$V_p$ probe characteristic. This is apparent from the experimental fact that the parallel electron velocity distribution function (EVDF) is almost Maxwellian. The values of $n_e$ and $T_e$ can be determined from the measured EVDF. The radial electric filed $E_r$ is obtained from the derivative of the plasma space potential with respect to r, -$dV_p/d_r$, where r is the radial position. The ion-neutral and electron-neutral collision mean free paths are around 1 m and 5 m in the experimental region, respectively. These values are much larger than the distance between the anode hole and the end plate. In the present experiment, the discharge current $I_d$ is kept constant at 10 A. The discharge voltage $V_d$ is then 110 V with the magnetic field strength $B_z$ = 0.19 T in the experimental region. The magnetic filed is measured point-by-point with small magnetic loop probe. The size of the magnetic probe is 10 mm in diameter and 20 mm in length.

Fig.1 Schematic diagram of the string plasma device (SPD). The profile of the axial magnetic field $B_z$ is shown in the upper frame.

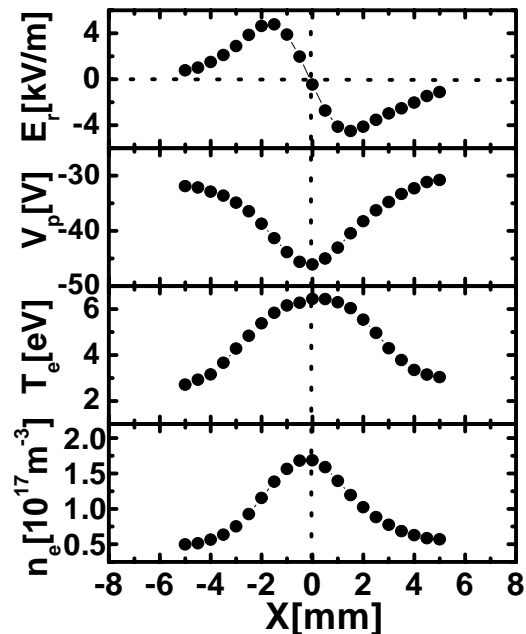

Fig.2 The radial profiles of the magnitude of the inwardly directed radial electric filed $E_r$, the plasma space potential $V_p$, the electron temperature $T_e$, and the electron density $n_e$ at the discharge current of 10 A.

3. Experimental Results

Figure 2 show the radial profiles of the magnitude of the inwardly directed radial electric filed $E_r$, the plasma space potential $V_p$, the electron temperature $T_e$, and the electron density $n_e$ at the discharge current of 10 A. The position y=0 mm corresponds to the center of the plasma cross section. The electron density has a hill-shaped profile with a half width of about 2.0 mm in the string plasma. Also, the string plasma has a steep electron temperature gradient over a plasma thickness of several millimeters: a hot plasma (~6.5 eV) in the central region and a cold plasma (0.5~1.0 eV) in the periphery region. The profile of the potential is axially symmetric and minimum at the center. The profile gradually increases as x became father from the center and then saturates in the outer region of the plasma, i.e., the radial potential profile is well type. The radial profiles of the magnitude of the inwardly directed radial electric field derived from $V_p$.

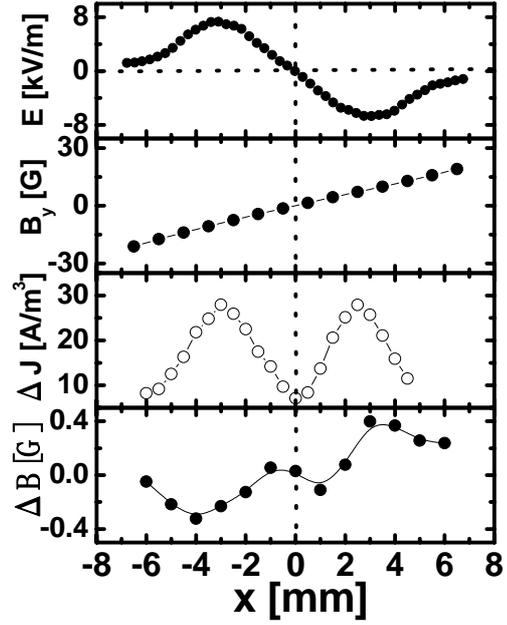

Fig.3 The radial profiles of the electric field $E_r$, the magnetic filed in vacuum, the net electron current to the axial direction, the displacement of the magnetic filed $B_\theta$.

Figure 3 show the radial profiles of the electric field $E_r$, the magnetic filed in vacuum $B_y$, the net electron current to the axial direction $\Delta J$ and the displacement of the magnetic filed $B_\theta$. The radial characteristic length $r_p$ of the plasma is considerably smaller than the ion Larmor radius. In our plasma, therefore, the plasma space potential $V_p$ varies in the radial direction in the range of $r_p$: $V_p$ is high in the outer region of the plasma and is minimum at the center. As a result, the inwardly directed radial electric field, $E_r$ naturally occurs and it also varies in the range of $r_p$. Because the ions can be considered unmagnetized under this condition, an axial current can arise in our plasma due to the $E_r \times B_\theta$ drift motion of the electrons. This hall current generates the magnetic field in an outside area of the plasma column resulting the inner magnetic field in the plasma column decrease and the outer magnetic field increase. The characteristics of the hall current indicates that the increase of the diamagnetic effects in a plasma.

4. Conclusion

The hall current in the periphery a filamentary plasma with the inwardly directed radial electric field, $E_r$, has been investigated experimentally by utilizing a TPD type dc discharge. In a steady-state magnetized filamentary plasma density gradient-scale length $L_n$ of the plasma column is smaller than the ion Larmor radius $\rho_i$ and much larger than the electron Larmor radius $\rho_e$, ($\rho_e \ll L_n < \rho_i$). This situation can be viewed that ions aren't trapped in a radial electric-potentail-well which is a result of the charge separation between an electron "string" confined along an axial magnetic field. When the multi-cusp magnetic filed $B_\theta$ is applied perpendicular to the stationary background magnetic field $B_z$ and the electric filed in the radial direction $E_r$, the hall current by the $E_r \times B_\theta$ drift of electrons is induced surrounding a magnetized filamentary plasma. This hall current generates the magnetic field in an outside area of the plasma column resulting the inner magnetic field in the plasma column decrease and the outer magnetic field increase. The characteristics of the hall current indicates that the increase of the diamagnetic effects in a plasma.

4